\documentclass[aps,prx,superscriptaddress,amsfonts,amsmath,amssymb,showpacs,floatfix,reprint]{revtex4-1}

\usepackage[english]{babel}
\usepackage{graphicx}					
\usepackage[pdftex]{epsfig}
\usepackage{ragged2e}
\usepackage[caption=false]{subfig}
\usepackage{amsmath,amssymb}
\usepackage{bbm}
\usepackage{bm}
\usepackage{times}
\usepackage[colorlinks,linkcolor=blue,citecolor=blue,urlcolor=blue]{hyperref}
\usepackage{color}
\usepackage[table,xcdraw]{xcolor}
\usepackage{epstopdf}
\usepackage{physics}
\usepackage[export]{adjustbox}
\usepackage{dsfont}
\usepackage{float}
\usepackage{makecell}

\newcommand{\be}{\begin{equation}}
\newcommand{\ee}{\end{equation}}

\newcommand{\bee}{\begin{equation*}}
\newcommand{\eee}{\end{equation*}}

\definecolor{cadmiumgreen}{rgb}{0.0, 0.42, 0.24}

\begin{document}

\title{$\mathbb{Z}_2$ vortices in the ground states of classical Kitaev-Heisenberg models}
\author{E. Seabrook}
\affiliation{Dahlem Center for Complex Quantum Systems and Fachbereich Physik, Freie Universit\"at Berlin, 14195 Berlin, Germany}
\author{M. L. Baez}
\affiliation{Max Planck Institute for the Physics of Complex Systems, N\"othnitzer Str. 38, 01187 Dresden, Germany}
\author{J. Reuther}
\affiliation{Dahlem Center for Complex Quantum Systems and Fachbereich Physik, Freie Universit\"at Berlin, 14195 Berlin, Germany}
\affiliation{Helmholtz-Zentrum Berlin f\"{u}r Materialien und Energie, Hahn-Meitner Platz 1, 14109 Berlin, Germany} 

\date{\today}
\pacs{}

\begin{abstract}
The classical nearest neighbor Kitaev-Heisenberg model on the triangular lattice is known to host $\mathds{Z}_2$ spin-vortices forming a crystalline superstructure in the ground state. The $\mathds{Z}_2$ vortices in this system can be understood as distortions of the local 120$^\circ$ N\'eel parent order of the Heisenberg-only Hamiltonian. Here, we explore possibilities of stabilizing further types of $\mathds{Z}_2$ vortex phases in Kitaev-Heisenberg models including those which rely on more complicated types of non-collinear parent orders such as tetrahedral states. We perform extensive scans through large classes of Kitaev-Heisenberg models on different lattices employing a two-step methodology which first involves a mean-field analysis followed by a stochastic iterative minimization approach. When allowing for longer-range Kitaev couplings we identify several new $\mathds{Z}_2$ vortex phases such as a state based on the 120$^\circ$ N\'eel order on the triangular lattice which shows a coexistence of different $\mathds{Z}_2$ vortex types. Furthermore, perturbing the tetrahedral order on the triangular lattice with a suitable combination of first and second neighbor Kitaev interactions we find that a kagome-like superstructure of $\mathds{Z}_2$ vortices may be stabilized where vortices feature a counter-rotating winding of spins on different sublattices. This last phase may also be extended to honeycomb lattices where it is related to cubic types of parent orders. In total, this analysis shows that $\mathds{Z}_2$ vortex phases appear in much wider contexts than the 120$^\circ$ N\'eel ordered systems previously studied.

\end{abstract}

\maketitle
\section{Introduction}
Topological defects are local perturbations of an otherwise homogeneous system which cannot be removed by any continuous operation. In condensed matter systems, topological defects have a long and active history of investigation where they appear in a colorful variety of different types including screw dislocations in crystals~\cite{nabarro52,ulvestad15}, magnetic skyrmions~\cite{rosner06,muhlbauer09,nagaosa13}, magnetic monopoles in spin ice~\cite{bramwell01,castelnovo08,morris09,jaubert11} and quantum vortices in superconductors and superfluids~\cite{essmann67,yarmchuk79,minnhagen87,brandt95,madison00,abo01}. A prototypical microscopic situation inducing topological defects arises when a two-dimensional system consists of local U(1) degrees of freedom such as the phase field of a superconducting film or the in-plane spin direction of an $XY$-magnet. In this case a vortex is formed if the U(1)-phase winds an integer number of times around the center of the perturbation, leading to a classification of defects in terms of a $\mathds{Z}$-quantized vorticity~\cite{minnhagen87,kenna06}. The insights about the thermodynamic properties of such phases, including vortex formation and binding/unbinding, as described in the seminal works by Berezinskii, Kosterlitz and Thouless (BKT)~\cite{berezinskii71,kosterlitz72,kosterlitz73} are of paramount importance in condensed matter physics.

Interestingly, a variant of the aforementioned $\mathds{Z}$-vortices also occurs in Heisenberg magnets if the system possesses a non-collinear local order parameter. This is, for example, realized in a nearest neighbor Heisenberg antiferromagnet on the triangular lattice which forms a 120$^\circ$ N\'eel ordered ground state~\cite{kawamura84,southern93,kawamura07,kawamura10}. Vortex excitations arising at finite temperatures then consist of deformations of this parent state where the local trio of spins performs a full rotation as one moves around the vortex core. Most importantly, these vortices are of $\mathds{Z}_2$-type~\cite{kawamura84} which implies that any pair of two vortices -- regardless of their precise microscopic realization -- can always be continuously transformed such that they mutually annihilate. In other words, there is no distinction between vortices and anti-vortices implying that they are topologically equivalent.

While the $\mathds{Z}_2$-vortex scenario is much less explored compared to the $\mathds{Z}$-case, there has recently been increasing interest in such phases as it has been realized~\cite{rousochatzakis16} that $\mathds{Z}_2$-vortices form stable defects in the ground states of classical triangular Heisenberg antiferromagnets when Kitaev interactions~\cite{kitaev06,jackeli09,chaloupka10,singh12,kimchi14} are added. The novel aspect of this observation is that the $\mathds{Z}_2$-vortices are not induced by thermal fluctuations as in the BKT transition but result from an interplay of frustration from isotropic Heisenberg and anisotropic Kitaev interactions. Particularly, already an infinitesimal Kitaev coupling is sufficient to generate a triangular crystalline superstructure of $\mathds{Z}_2$-vortices where the distance between vortices decreases with increasing Kitaev interaction. In momentum space, the vortex crystal formation manifests in a characteristic shift of the 120$^\circ$ magnetic Bragg peak away from the corners of the first Brillouin zone, accompanied by the emergence of subleading satellite peaks. The initial observation of $\mathds{Z}_2$-vortices in the ground states of classical triangular Kitaev-Heisenberg models~\cite{rousochatzakis16} motivated a series of follow-up works where the deformation of the 120$^\circ$ N\'eel state into a vortex crystal has been investigated for longer-range Heisenberg couplings~\cite{yao16,yao18}, Dzyaloshinskii-Moriya interactions~\cite{osorio19}, honeycomb lattices~\cite{yao16,yao18} (or an interpolation between triangular- and honeycomb lattices~\cite{kishimoto18}), quantum spins~\cite{shinjo16,kos17} (including their dynamics in a semiclassical approximation~\cite{li19}), and from a material perspective~\cite{becker15,catuneanu15}.

In general, the allowed types of topological defects in two-dimensional systems are determined by the first homotopy group $\pi_1$ of the system's order parameter space~\cite{kenna06}. For the XY-magnet and the triangular Heisenberg antiferromagnet the order parameter spaces are U(1) and SO(3), respectively, such that the aforementioned nature of their vortices follows from the properties $\pi_1(\text{U(1)})=\mathds{Z}$ and $\pi_1(\text{SO(3)})=\mathds{Z}_2$. Particularly, since any non-collinear ordered isotropic spin system has (at least) an SO(3) order parameter space, one may expect that $\mathds{Z}_2$-vortices do not only occur for 120$^\circ$-N\'eel ordered parent states but also appear as deformations of any other coplanar or non-coplanar parent state, also including those orders where the magnetic unit cell consists of more than three sites. [Note that collinear ordered isotropic spin systems must be excluded since the first homotopy group of their order parameter space S$_2$ is $\pi_1(\text{S}_2)=0$ such that they cannot host topological defects]. The possibility of stabilizing $\mathds{Z}_2$-vortices in these generalized magnetic environments is, however, largely unexplored so far.

In this paper, we investigate topological defects in Kitaev-Heisenberg models from a more general viewpoint by addressing the question of which other $\mathds{Z}_2$-vortex phases can occur, besides their emergence out of 120$^\circ$-N\'eel order in triangular lattice Kitaev-Heisenberg models. To this end, we particularly focus on parent states with four-sublattice tetrahedral and eight-sublattice cubic types of orders which occur on triangular and honeycomb lattices with longer-range Heisenberg interactions, and then perturb the systems by adding Kitaev couplings. We note that perturbations of Kitaev-type are well-suited to explore such phenomena since they induce a non-trivial anisotropic frustration mechanism as needed for the formation of $\mathds{Z}_2$-vortices but do not enforce any easy-plane anisotropy (which may quickly result in the more conventional $\mathds{Z}$-type vortex formation as in $XY$-magnets~\cite{cuccoli03,cuccoli03_2}). Our study involves extensive scans through a wide range of classical spin Hamiltonians exhibiting longer-range Heisenberg interactions (to tune the systems to the desired parent states) and longer-range Kitaev couplings (to generate $\mathds{Z}_2$-vortex phases). For an efficient survey, we pursue a two-step strategy: We first employ a faster (but approximative) mean-field scheme~\cite{reimers92} to identify candidate systems based on the characteristic shift of magnetic Bragg peaks associated with the onset of $\mathds{Z}_2$-vortices. This reduced number of candidate models is then treated with a stochastic iterative minimization method~\cite{Walker1980,Sklan2013} to find the real ground states.

Our main results can be summarized as follows: After briefly revisiting the $\mathds{Z}_2$-vortex crystal in the triangular lattice Kitaev-Heisenberg model as first investigated in Ref.~\cite{rousochatzakis16}, we identify another type of vortex phase in this system which emerges out of the 120$^\circ$-N\'eel order upon adding second neighbor Kitaev interactions. Apart from a different arrangement of Bragg peaks in momentum space, this phase is characterized by the coexistence of two types of $\mathds{Z}_2$-vortices exhibiting different (albeit topologically identical) rotation axes of the local tripods of spins. We then tune the parent Heisenberg system into the regime $1/8 < J_2/J_1 < 1$, where $J_1$ ($J_2$) is the first (second) neighbor antiferromagnetic interaction on the triangular lattice. For these couplings the ground state is degenerate, supporting any type of magnetic order where the sum of all spins in a four-site magnetic unit cell vanishes~\cite{chubukov92}. This degeneracy enables a large variety of possible deformations and is, hence, a particularly interesting starting point for investigating $\mathds{Z}_2$-vortex phases. We find that upon adding a suitable combination of longer-range Kitaev couplings, the system realizes a kagome superstructure of $\mathds{Z}_2$-vortices. In each individual vortex the four sites of the magnetic unit cell split up into pairs which perform a counterrotating motion through a manifold of tetrahedral states as one moves around the vortex core. Finally, we identify a doubled version of this phase for the Kitaev-Heisenberg model on the honeycomb lattice where the eight-site magnetic unit cell of degenerate cubic-type magnetic orders splits up into groups of four sites which again perform a collective counterrotation in each vortex. In total, this analysis highlights the richness of physical phenomena in Kitaev-type magnets and opens the door to more refined investigations of the identified phases.

The rest of the paper is organized as follows: In Sec.~\ref{parent_ham} we first introduce and discuss the Heisenberg parent Hamiltonians and their non-collinear ordered ground states. These are the 120$^\circ$-N\'eel state on the triangular lattice (Sec.~\ref{120neel}), the tetrahedral states on the triangular lattice (Sec.~\ref{tetrahed}) and the cubic states on the honeycomb lattice (Sec.~\ref{cubic}). The following Sec.~\ref{methods} introduces the two methods which are used to treat these systems when adding Kitaev interactions: In Sec.~\ref{meanfield} we present a mean-field scheme for the susceptibility while Sec.~\ref{im} discusses the iterative minimization technique. We then present the result of these two approaches in Sec.~\ref{results} where Sec.~\ref{mf_results} first applies the mean-field method to identify possible candidate vortex phases. The momentum-resolved mean-field responses of the resulting four models are briefly discussed. In the following Sec.~\ref{im_results} we treat these models one by one with iterative minimization and describe in detail the spin arrangements in the identified vortex phases. The paper ends with a conclusion in Sec.~\ref{DandC}.

\begin{figure}
\includegraphics[width=0.99\linewidth]{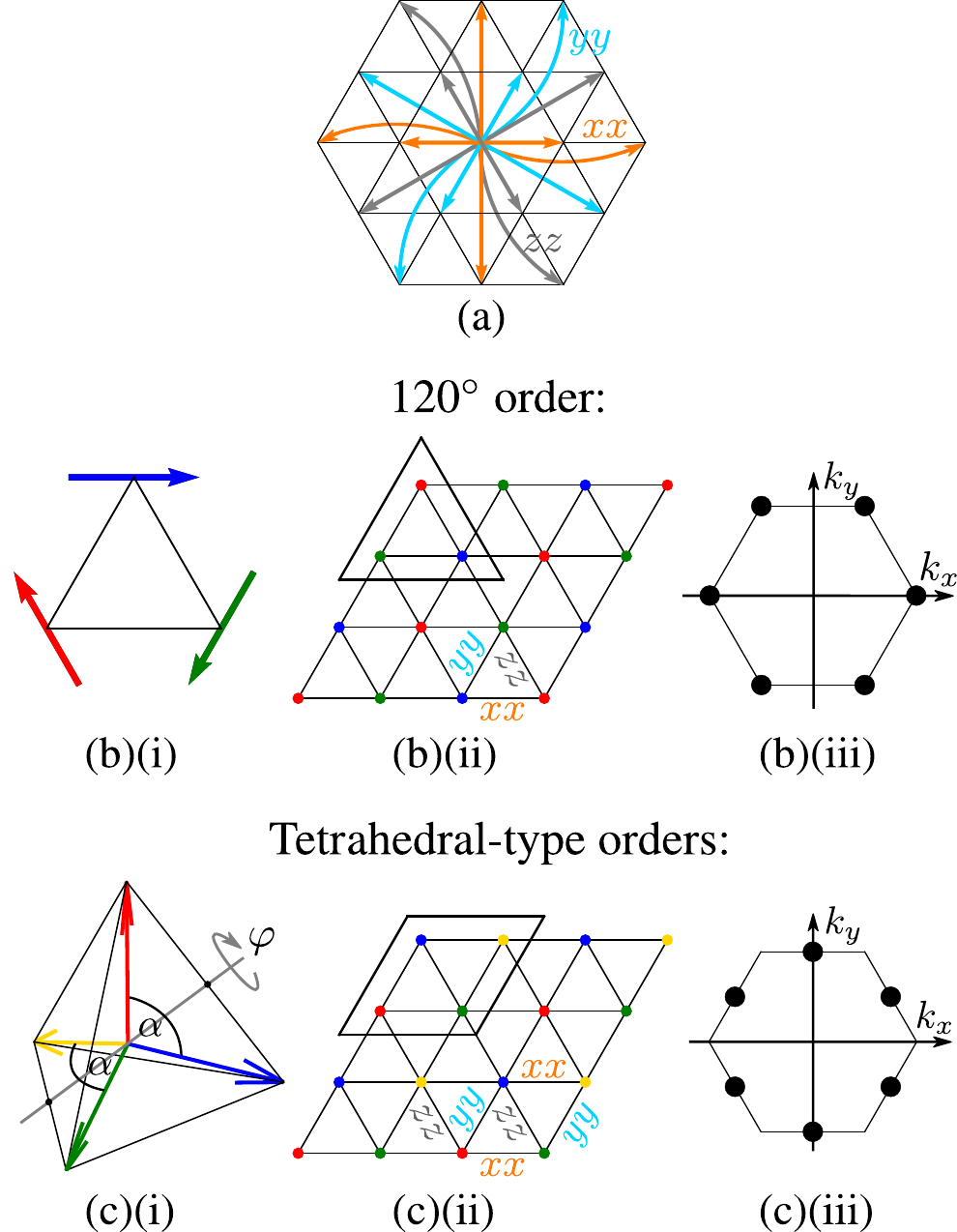}
\caption{\label{fig1}The triangular lattice and its commensurate non-collinear types of order. (a) First, second, and third neighbor bonds on the triangular lattice. Orange, cyan, and gray arrows represent Kitaev bonds of $xx$, $yy$ and $zz$-type, respectively. (b) The three-sublattice $120^\circ$-N\'eel state. (b)(i) depicts the three spin orientations defining this state and (b)(ii) shows their spatial arrangement in the lattice with selected nearest neighbor Kitaev-bonds labelled `$xx$', $\ldots$. (b)(iii) indicates the magnetic Bragg-peak location of the $120^\circ$-N\'eel state in the first Brillouin zone (black hexagon). (c) The four-sublattice tetrahedral-type orders. (c)(i) illustrates the construction of all degenerate spin orders in this phase (see text for details) while (c)(ii) displays the arrangement of the sublattices with selected nearest neighbor Kitaev-bonds labelled `$xx$', $\ldots$. (c)(iii) indicates the momentum-space location of these orders in the first Brillouin zone (black hexagon).}
\end{figure}
\begin{figure}
\includegraphics[width=0.9\linewidth]{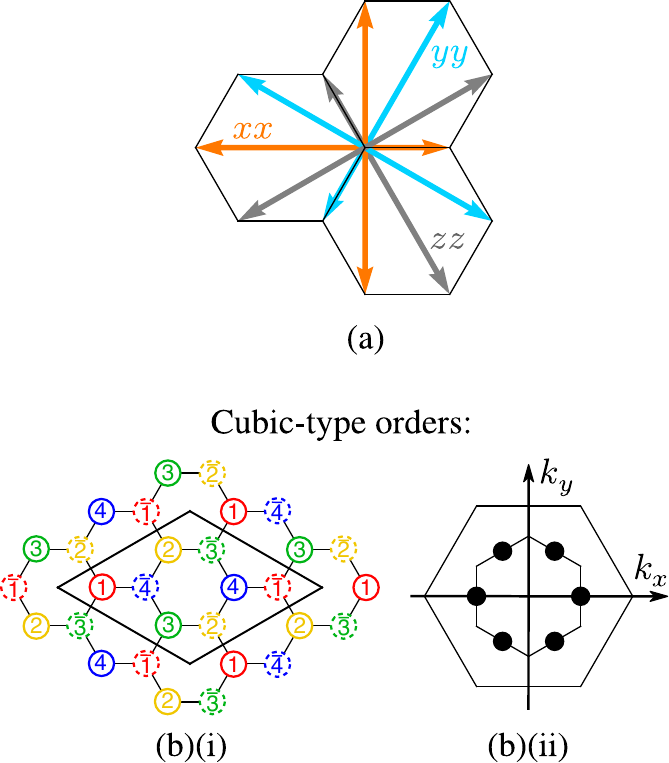}
\caption{\label{fig2}(a) First, second, and third neighbor bonds on the honeycomb lattice where orange, cyan, and gray arrows represent Kitaev bonds of $xx$, $yy$ and $zz$-type, respectively. (b) The magnetic unit cell of the cubic states has eight sublattices denoted $1$, $2$, $3$, $4$ (shown as full circles) and $\bar{1}$, $\bar{2}$, $\bar{3}$, $\bar{4}$ (shown as dashed circles). In each of these two sets, the spins $\mathbf{S}_\alpha$ and $\mathbf{S}_{\bar{\alpha}}$ sum up to zero and $\mathbf{S}_{\alpha}=-\mathbf{S}_{\bar{\alpha}}$ for $\alpha\in\{1,2,3,4\}$. (c) Momentum-space location of the cubic order where the inner (outer) hexagon corresponds to the first (extended) Brillouin zone.}
\end{figure}

\section{Parent Heisenberg Hamiltonians}\label{parent_ham}
In this section, we specify the class of parent Heisenberg Hamiltonians $H_0$ for which we will investigate $\mathds{Z}_2$ vortex formation. The Hamiltonians $H_0$ take the form
\begin{equation}
H_0= \sum_n J_n\sum_{\langle ij\rangle_n}\mathbf{S}_i\mathbf{S}_j\;,\label{ham}
\end{equation}
where $\langle i j\rangle_n$ denotes pairs of $n$th neighbor sites on the respective lattice [see Figs.~\ref{fig1}(a) and \ref{fig2}(a)] and $J_n$ is the corresponding coupling strength (in the following, we will restrict ourselves to interactions up to third neighbors). Furthermore, $\mathbf{S}_i$ is a normalized three-component vector representing the classical spin on site $i$. Below, we will introduce and discuss the non-collinear ordered ground states of Eq.~(\ref{ham}) which may potentially host $\mathds{Z}_2$ vortices upon adding Kitaev couplings. For the triangular lattice Heisenberg model, we will focus on three-sublattice $120^\circ$-N\'eel order and four-sublattice tetrahedral-type magnetic orders. Apart from incommensurate spin spirals, these two states already cover all possible non-collinear ordered phases in the triangular lattice Heisenberg model~\cite{messio11} (note that incommensurate spiral phases are ignored here since their unbounded unit cells do not support the types of topological defects discussed here). Furthermore, for the honeycomb Heisenberg model we will consider eight-sublattice cubic spin phases.

\subsection{$120^\circ$-N\'eel order on the triangular lattice}\label{120neel}
The 120$^\circ$-N\'eel order is a coplanar state including three sublattices oriented at 120$^\circ$ with respect to one another [Fig.~\ref{fig1}(b)(i)], creating a three-site magnetic unit cell [Fig.~\ref{fig1}(b)(ii)]. The sublattices are arranged such that each pair of second neighbor spins has the same direction. This type of order represents the ground state in the case of only nearest neighbor antiferromagnetic Heisenberg couplings but remains intact for all second neighbor couplings $J_2<J_1/8$~\cite{jolicoeur90,chubukov92,messio11}. As discussed below, the addition of second neighbor couplings $J_2$ turns out beneficial for the formation of a vortex crystal from 120$^\circ$-N\'eel order. We will, hence, consider the triangular lattice Heisenberg model with $J_1=1$ and $J_2=-1$. Note that due to the spin-isotropy of Eq.~(\ref{ham}), any rotated version of the 120$^\circ$-N\'eel order is also a ground state of the system which, particularly, implies that the plane which is common to all spins is arbitrary. The magnetic Bragg peaks describing this order are located at the corners of the first Brillouin zone, as can be seen in Fig.~\ref{fig1}(b)(iii).

\subsection{Tetrahedral orders on the triangular lattice}\label{tetrahed}
The triangular lattice Heisenberg model hosts another commensurate non-collinear phase which is characterized by a four-site magnetic unit cell as depicted in Fig.~\ref{fig1}(c)(ii). It occurs in the presence of antiferromagnetic $J_1$ and $J_2$ couplings when $1/8 < J_2/J_1 < 1$~\cite{jolicoeur90,chubukov92,messio11}. As a representative point in this phase we will consider the case $J_1=1$ and $J_2=0.5$ below. Most strikingly, this state is degenerate and admits all configurations where the sum of the spins in each unit cell vanishes,
\begin{equation} 
\mathbf{S}_1+\mathbf{S}_2+\mathbf{S}_3+\mathbf{S}_4=0\;.\label{degenerate}
\end{equation}
The manifold of states obeying this condition may be illustrated by starting with the tetrahedral state which is a particularly symmetric representative of the degenerate spin configurations. As shown in Fig.~\ref{fig1}(c)(i), the four spin directions defining the tetrahedral state are given by the vectors from the center of a tetrahedron to its corners, which form an angle of $109.5^\circ$ with respect to one another. Up to global spin rotations of all four sublattices, the other degenerate states may now be constructed as follows: We first define a rotation axis which connects the midpoints of two opposite edges of the tetrahedron [see gray line in Fig.~\ref{fig1}(c)(i)]. This axis also includes the center of the tetrahedron. Next consider two spins whose endpoints are connected by one of these edges such as, e.g., the blue and red spins in Fig.~\ref{fig1}(c)(i). All configurations obtained by rotating these two spins by an angle $\varphi$ around the specified axis, while keeping the other pair of spins fixed, obey Eq.~(\ref{degenerate}). Additionally, to fully cover the degenerate manifold of states one needs to allow for variations of the angle $\alpha$ between these pairs of spins (this last deformation must be performed such that the rotation axis remains symmetrically centered between the two spins of each pair). Apart from overall rotations of all four sublattices, the two angles $\varphi$ and $\alpha$ allow one to access all degenerate states. 

It is clear that the degenerate states are, generically, non-planar, however, for special angles $\phi$ and $\alpha$ one may also obtain planar or even collinear states (note that quantum fluctuations in a semiclassical $1/S$ approximation induce an order-by-disorder mechanism which energetically prefers collinear states with $\alpha=0$~\cite{jolicoeur90,chubukov92}). Since the spins in each of these degenerate states point towards the corners of a deformed tetrahedron, we denote them as `tetrahedral-type' orders.

Various different types of vortices seem possible when the tetrahedral phase is perturbed by Kitaev couplings. Firstly, a vortex may not make use of the system's degeneracy. In this case, the perturbation would pick out a certain state with fixed angles $\varphi$, $\alpha$. The deformation of this state as one moves around a vortex core would then only correspond to a {\it simultaneous} rotation of all four spins around a certain axis. Secondly, the angles $\varphi$ and/or $\alpha$ may change along a loop around a vortex such that the local spin configurations cover parts of the degenerate manifold. As we will see below, longer-range Kitaev couplings realize the second possibility. Particularly, two pairs of spins [e.g. (blue, red) and (yellow, green) in Fig.~\ref{fig1}(c)(i)] perform a counterrotating motion with angles $\varphi$ and $-\varphi$, respectively.

We finally mention that the magnetic Bragg peaks corresponding to each of these degenerate states are located at the midpoints of the edges of the first Brillouin-zone, as illustrated in Fig.~\ref{fig1}(c)(iii).

\subsection{Cubic orders on the honeycomb lattice}\label{cubic}
In addition to the triangular lattice we also investigate vortex formation on the honeycomb lattice representing the simplest non-Bravais lattice in two dimensions. The honeycomb lattice Heisenberg model features two commensurate non-collinear ground state orders, a four-sublattice tetrahedral phase, similar to the one in the previous subsection, and an eight-sublattice cubic phase~\cite{messio11}. Since the tetrahedral-type states did not show any ground-state vortices upon adding Kitaev interactions we will not further discuss this phase in the following. The remaining cubic phase has a rather large extent in the $J_1$-$J_2$-$J_3$ parameter space and occurs for ferromagnetic and antiferromagnetic $J_1$ if $J_2>0$ and $J_3>0$ are sufficiently large. In the case of ferromagnetic $J_1<0$, the cubic order can even be stabilized without $J_2$ couplings~\cite{messio11,rastelli79,fouet01}. This regime will be investigated below where we choose the specific coupling strengths $J_1=-1$ and $J_3=1$.

The cubic order is degenerate and can be considered as two copies of the tetrahedral states of the previous subsection~\cite{fouet01}. This property results from the bipartite nature of the honeycomb lattice which, by itself, consists of two triangular sublattices, here denoted $A$ and $B$. In the presence of cubic order, the unit cell is enlarged, consisting of four sublattices of type $A$ (labelled $1$, $2$, $3$, $4$ in the following) and four sublattices of type $B$ (labelled $\bar{1}$, $\bar{2}$, $\bar{3}$, $\bar{4}$), see Fig.~\ref{fig2}(b)(i) for the precise definition of the sublattices. All properties of the tetrahedral state discussed above, such as the conditions on the spin sums, 
\begin{equation}
\sum_{\alpha=1}^4 \mathbf{S}_\alpha=0\quad\text{and}\quad\sum_{\alpha=1}^4 \mathbf{S}_{\bar{\alpha}}=0\;,
\end{equation}  
remain separately valid for both sets of sites. Furthermore, the spin orientations on the sublattices $A$ and $B$ are antiparallel, i.e., $\mathbf{S}_{\alpha}=-\mathbf{S}_{\bar{\alpha}}$ for $\alpha\in\{1,2,3,4\}$. The term `cubic' refers to the fact that two copies of the ideal tetrahedral order depicted in Fig.~\ref{fig1}(b)(i) (where one copy has reversed spin directions) yields a state with spins pointing at the corners of a cube when plotted with the same origin. As discussed in more detail below, this phase also hosts $\mathds{Z}_2$-vortices where in each sublattice $A$ and $B$, pairs of spins perform a counterrotating deformation defined by angles $\varphi$ and $-\varphi$.

In momentum space the cubic orders reside at the midpoints of the edges of the first Brillouin zone, see Fig.~\ref{fig2}(b)(ii).

\section{Methods}\label{methods}
All types of magnetic orders presented in Sec.~\ref{parent_ham} will be perturbed by Kitaev interactions of the form
\begin{equation}
H_K=\sum_n K_n\sum_{\langle ij\rangle_n \in \gamma}S^{\gamma}_i S^{\gamma}_j\;,\label{ham_kitaev}
\end{equation} 
where $\gamma\in\{x,y,z\}$ denotes a spin component specific to the corresponding $n$-th neighbor bond $\langle ij\rangle_n$, as illustrated in Figs.~\ref{fig1}(a) and \ref{fig2}(a). In contrast to $H_0$, the ground state of the full anisotropic Hamiltonian $H=H_0+H_K$ is no longer solvable in closed analytical form, even in the classical case. One, therefore, either has to rely on approximate analytical methods or on numerical techniques. Here, we employ a combination of both, where we first use a mean-field scheme (see e.g. Ref.~\cite{reimers92} for a similar approach) to reduce the full class of Kitaev-Heisenberg models to a smaller set of a few candidate systems. These models are then treated within a numerical iterative minimization scheme~\cite{Walker1980,Sklan2013} to investigate the ground state spin configurations and to possibly identify vortex formation. Here, we briefly describe the two employed methods.

\subsection{Mean-field theory for the magnetic susceptibility}\label{meanfield}
For notational convenience, we combine the Heisenberg and Kitaev interactions $J_n$ and $K_n$ into a single coupling $J_{ij}^\mu$ such that the full Hamiltonian $H=H_0+H_K$ reads
\begin{equation}
H=\sum_\mu\sum_{(ij)}J_{ij}^\mu S_i^\mu S_j^\mu\;,\label{ham_j}
\end{equation}
where $(i,j)$ denotes pairs of sites (which are summed over only once). Our mean-field approach relies on the standard decoupling of quadratic spin interactions,
\begin{equation}
S_i^\mu S_j^\mu\rightarrow S_i^\mu\langle S_j^\mu\rangle+\langle S_i^\mu\rangle S_j^\mu -\langle S_i^\mu\rangle\langle S_j^\mu\rangle\;,
\end{equation}
which yields a self-consistent condition for the thermal spin-expectation values $\langle S_i^\mu\rangle$,
\begin{equation}
\langle S_i^\mu \rangle=\frac{1}{2}\tanh\left(\frac{B_i^\mu-\sum_j J_{ij}^\mu \langle S_j^\mu \rangle}{2 k_\text{B}T}\right)\;.\label{self_consistent_s}
\end{equation}
Here, we temporarily allow for local magnetic source fields $B_i^\mu$ which are added via $H\rightarrow H-\sum_i\mathbf{B}_i\mathbf{S}_i$. For methodological reasons to become clear below, we also consider finite temperatures $T$. 

One may now define a local, zero-field susceptibility via
\begin{equation}
\chi_{ij}^{\mu\mu}=\frac{\partial \langle S_i^\mu\rangle}{\partial B_j^\mu}\Bigg|_{B_j\rightarrow 0}\;.\label{local_sus}
\end{equation}
Exploiting Eq.~(\ref{self_consistent_s}) and restricting to the paramagnetic phase where all spin-expectation values $\langle S_i^\mu\rangle$ vanish, one obtains a self-consistent equation for $\chi^{\mu\mu}_{ij}$,
\begin{equation}
\chi_{ij}^{\mu\mu}=\frac{1}{4k_\text{B} T}\left(\delta_{ij}-\sum_l J_{il}^\mu\chi_{lj}^{\mu\mu}\right)\;.\label{self_consistent_chi}
\end{equation}
Since all models investigated here only consist of diagonal couplings in spin space, different components $\mu$ do not mix in this equation. Assuming a lattice with an $N_a$-atomic unit cell (note that $N_a$ refers to the unit cell of the lattice, not to the magnetic unit cell), Eq.~(\ref{self_consistent_chi}) can be straightforwardly solved by Fourier-transforming $\chi_{ij}^{\mu\mu}$ via
\begin{equation}
\tilde{\chi}_{\rho\sigma}^{\mu\mu}(\mathbf{k})=\sum_{\Delta\mathbf{R}_{ab}}e^{-i\mathbf{k}\Delta \mathbf{R}_{ab}}\chi_{a\rho b\sigma}^{\mu\mu}\;,
\label{fourier}
\end{equation}
and equivalently for $\tilde{J}_{\rho\sigma}^\mu(\mathbf{k})$. Here, we have decomposed the site index $i$ into $i\rightarrow\{a,\rho\}$ where $a$ denotes the unit cell of site $i$ and $\rho\in\{1,\ldots,N_a\}$ is a sublattice index (in the same way $j\rightarrow\{b,\sigma\}$). Furthermore, the site positions $\mathbf{r}_i$ are split up into $\mathbf{r}_i=\mathbf{R}_a+\boldsymbol{\xi}_\sigma$ where $\mathbf{R}_a$ are unit-cell positions and $\boldsymbol{\xi}_\sigma$ denotes displacements within the unit cells. Distance vectors between unit cells are written as $\Delta \mathbf{R}_{ab}=\mathbf{R}_a-\mathbf{R}_b$. With these definitions, the solution of Eq.~(\ref{self_consistent_chi}) is given by
\begin{equation}
\tilde{\chi}^{\mu\mu}(\mathbf{k})=[4k_\text{B}T\mathds{1}+\tilde{J}^\mu(\mathbf{k})]^{-1}\;,\label{solution}
\end{equation}
where we have dropped the sublattice indices $\rho$, $\sigma$ in $\tilde{\chi}^{\mu\mu}(\mathbf{k})$ and $\tilde{J}^\mu(\mathbf{k})$ to indicate that these quantities can be interpreted as $N_a\times N_a$-matrices in sublattice space. Furthermore, $\mathds{1}$ denotes the $N_a\times N_a$ identity matrix and the exponent $-1$ stands for the usual matrix inversion.

The matrix-valued susceptibility $\tilde{\chi}^{\mu\mu}(\mathbf{k})$ is closely related to the usual scalar momentum-resolved susceptibility $\chi^{\mu\mu}(\mathbf{k})$, defined by
\begin{equation}
\chi^{\mu\mu}(\mathbf{k})=\frac{1}{N}\frac{\partial \langle S^\mu(\mathbf{k})\rangle}{\partial B^\mu(\mathbf{k})}\Bigg|_{\{B_i\}\rightarrow0}
\end{equation}
where
\begin{equation}
\langle S^\mu(\mathbf{k})\rangle=\sum_i e^{-i\mathbf{k}\mathbf{r}_i}\langle S_i^\mu\rangle
\end{equation}
and equivalently for $B^\mu(\mathbf{k})$. Using Eqs.~(\ref{local_sus}) and (\ref{fourier}) one finds
\begin{equation}
\chi^{\mu\mu}(\mathbf{k})=\frac{1}{N_a}\sum_{\rho,\sigma} e^{-i\mathbf{k}(\boldsymbol{\xi}_\rho-\boldsymbol{\xi}_\sigma)}\tilde{\chi}^{\mu\mu}_{\rho\sigma}(\mathbf{k})\;.\label{chi_final}
\end{equation}

The investigations in the next section are based on the momentum-resolved susceptibility $\chi^{\mu\mu}(\mathbf{k})$, determined from Eqs.~(\ref{solution}) and (\ref{chi_final}). Starting at a sufficiently large initial $T$ and for a given set of couplings $J_n$ and $K_n$, the temperature is lowered until the right-hand side of Eq.~(\ref{solution}) becomes singular at certain momenta $\mathbf{k}$. At this critical mean-field temperature $T_\text{c}$, sharp peaks appear in the susceptibility $\chi^{\mu\mu}(\mathbf{k})$ signaling the onset of magnetic long-range order, while for $T<T_\text{c}$ the solution in Eq.~(\ref{solution}) is no longer valid. The peak positions in momentum space right at the critical temperature provide an approximation for the ground state magnetic order. Despite its mean-field character, this approach proves to be a simple but efficient tool for a first examination of the system's magnetic properties.

It is worth emphasizing that the resulting magnetic wave vectors coincide with those from a Luttinger-Tisza analysis~\cite{luttinger46,luttinger51,baez17}. An additional benefit of our approach is that the susceptibility $\chi^{\mu\mu}(\mathbf{k})$ contains the full momentum-resolved intensity distribution of the magnetic response which also allows one to identify subleading ordering tendencies. 

\subsection{Iterative minimization}\label{im}
In addition to the approximative mean-field approach of the last subsection we apply a numerical iterative minimization scheme~\cite{Walker1980,Sklan2013} which, up to statistical errors, finds the exact spin configuration of an energy minimum (which may, however, be local, see comments below). The motivation behind this method is based on the fact that in any genuine classical ground state, each spin must be aligned with its local field $\mathbf{h}_i$ whose components are defined by
\begin{equation}
h_i^\mu=-\frac{\delta H}{\delta S_i^\mu}=-\sum_{j}J_{ij}^\mu S_j^\mu\;,
\label{localfield}
\end{equation}
where in the last step we have assumed a Hamiltonian of the form of Eq.~(\ref{ham_j}).

The calculation starts with an initial state of $N$ spins randomly oriented. We then perform successive sweeps over the lattice. In each sweep we select $N$ spins at random (with repetitions allowed) which are successively oriented along its local field,
\begin{equation}
\mathbf{S}_i\longrightarrow \frac{\mathbf{h}_i}{|\mathbf{h}_i|}\;.
\label{reorient}
\end{equation}
This process thus generates a new configuration which, in each sweep, lowers the classical energy. Sweeps are performed until the energy difference between the new and old spin configuration falls below a predefined threshold ($10^{-13}$ in our case). For the triangular and honeycomb lattices we use system sizes $N=2611$ and $N=2814$ respectively, and implement open boundary conditions, unless stated otherwise.

Please note that iterative minimization is a stochastic method based on a steepest decent algorithm, and as such is prone to detect local minima instead of global ones. To overcome this problem, we run the algorithm a given number of times (10 - 20 times) starting from different random configurations, and choose the spin configuration that has the lowest energy.

\section{Results}\label{results}

\subsection{Mean-field results}\label{mf_results}
The models $H=H_0+H_\text{K}$ [see Eqs.~(\ref{ham}) and (\ref{ham_kitaev})], where $H_0$ denotes the parents Hamiltonians from Sec.~\ref{parent_ham} exhibiting 120$^\circ$-N\'eel, tetrahedral- and cubic types of orders, are first treated with the mean-field approach from Sec.~\ref{meanfield}. This approximative method allows us to efficiently sweep through large parameter regions of Kitaev couplings, hence, identifying possible candidate $\mathds{Z}_2$-vortex phases which are then investigated in more detail using the iterative minimization scheme. Even with this mean-field approach, the space of nearest neighbor up to third neighbor Kitaev couplings $K_1$, $K_2$, $K_3$, including positive and negative signs of each of them, is too large to be mapped out as a whole. Therefore, in the three-dimensional space $(K_1,K_2,K_3)$ of Hamiltonians $H_\text{K}$ we choose to explore nine different cuts including the three lines along the vectors $(1,0,0)$, $(0,1,0)$, $(0,0,1)$ (i.e., where only one Kitaev coupling is finite) and the six lines $(1,\pm1,0)$, $(1,0,\pm1)$, $(0,1,\pm1)$ (i.e., the diagonals in each plane where one Kitaev coupling vanishes). While this already gives a good coverage of the full phase diagram, an extended investigation, which we leave for future studies, could also include additional cuts and/or further types of interactions such as off-diagonal Dzyaloshinskii-Moriya couplings and $\Gamma$-exchange.

Our main diagnostic tool in this section is the peak structure of the magnetic susceptibility $\chi^{\mu\mu}(\mathbf{k})$ from Eq.~(\ref{chi_final}) right above the critical temperature $T_\text{c}$. While all the orders discussed in Sec.~\ref{parent_ham} reside at high symmetry points in momentum space, the formation of a vortex superstructure due to finite perturbations $H_\text{K}$ induces an additional (and typically incommensurate) periodicity of the real-space spin pattern, which in reciprocal space manifests in a continuous shift of the magnetic Bragg peaks away from these commensurate points~\cite{rousochatzakis16}. Our investigation below aims at identifying such characteristic shifts upon increasing the Kitaev couplings $K_n$. It is important to emphasize that, in general, this approach might be subject to two types of errors. Firstly, an existing vortex phase could be missed because the mean-field approach might be unable to resolve the peak shifting due to its approximate character. Secondly, an observed peak migration does not necessarily signal a vortex phase but might also indicate any other type of incommensurate spin configuration such as, e.g., a spin spiral. We, indeed, identified some models where this is the case, which will, however, not be further explored here. Instead, we concentrate on those phases which eventually show vortices with iterative minimization.

As discussed in more detail below, we identified four $\mathds{Z}_2$-vortex phases via the aforementioned shift of Bragg peaks, where two of them are based on the 120$^\circ$-N\'eel order (one is the well-known vortex phase first studied in Ref.~\onlinecite{rousochatzakis16}) while the other two are deformations of the tetrahedral and cubic orders, respectively. In Table~\ref{table:1}, we summarize the relevant information of these models where we specify the lattice, the couplings $J_1$, $J_2$, $J_3$ of the parent Hamiltonians $H_0$, including its ground state order, and the direction in parameter space $(K_1,K_2,K_3)$ along which a continuous Bragg peak shift is observed. In the following, we discuss the momentum-space structure of the mean-field susceptibility for each of these models. 
\begin{table}
\centering
\begin{tabular}{ |c|c|c|c|c|c|c| }
 \hline
 \textbf{Model} & \textbf{Lattice} & \makecell{\textbf{Heisenberg}\\ \textbf{order}} & $\bm{J_1}$ & $\bm{J_2}$ & $\bm{J_3}$ & $\bm{(K_1,K_2,K_3)}$\\ 
\hline\hline
 I & Triangular & 120$^\circ$-N\'eel & $1$ & $-1$ & $0$ & $(\pm1,0,0)$\\ 
\hline
 II & Triangular & 120$^\circ$-N\'eel & $1$ & $-1$ & $0$ & $(0,1,0)$\\ 
\hline
 III & Triangular & Tetrahedral & $1$ & $0.5$ & $0$ & $(-1,1,0)$\\
 \hline
 IV & Honeycomb & Cubic & $-1$ & $0$ & $1$ & $(0,-1,-1)$\\
\hline
\end{tabular}
\caption{The four candidate models for $\mathbb{Z}_2$-vortex phases found within mean-field theory. The columns specify the lattice, the underlying commensurate ground state order, the couplings $J_1$, $J_2$, $J_3$ of the parent Heisenberg Hamiltonian $H_0$ (which are held fixed in each model), and the direction $(K_1,K_2,K_3)$ in the space of Kitaev couplings along which a continuous Bragg peak shift is found.}
\label{table:1}
\end{table} 

\begin{figure*}
\centering
\includegraphics[width=\textwidth]{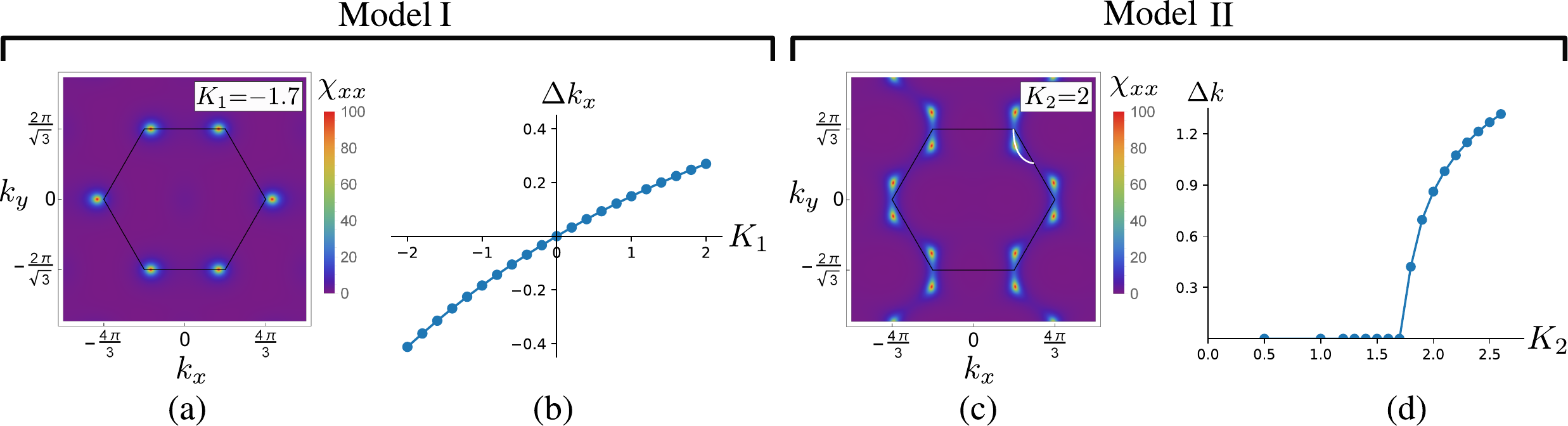}
\caption{Mean-field susceptibility $\chi^{xx}(\mathbf{k})$ and magnetic Bragg-peak evolution for models I and II on the triangular lattice: (a) Susceptibility $\chi^{xx}(\mathbf{k})$ of model I at $K_1=-1.7$, (b) peak displacement $\Delta k_x$ relative to the 120$^\circ$-N\'eel order position (corners of the first Brillouin zone) as a function of $K_1$ for model I, (c) susceptibility $\chi^{xx}(\mathbf{k})$ of model II at $K_2=2$. The white line shows the trajectory of the peak in the first quadrant while all other trajectories follow by symmetry.  (d) peak displacement $\Delta k=\sqrt{\Delta k_x^2+\Delta k_y^2}$ relative to the 120$^\circ$-N\'eel order position as a function of $K_2$ for model II. All data is collected right above the respective critical temperature $T_\text{c}$.}
\label{fig: 120MFT}
\end{figure*}

\begin{figure*}
\centering
\includegraphics[width=\textwidth]{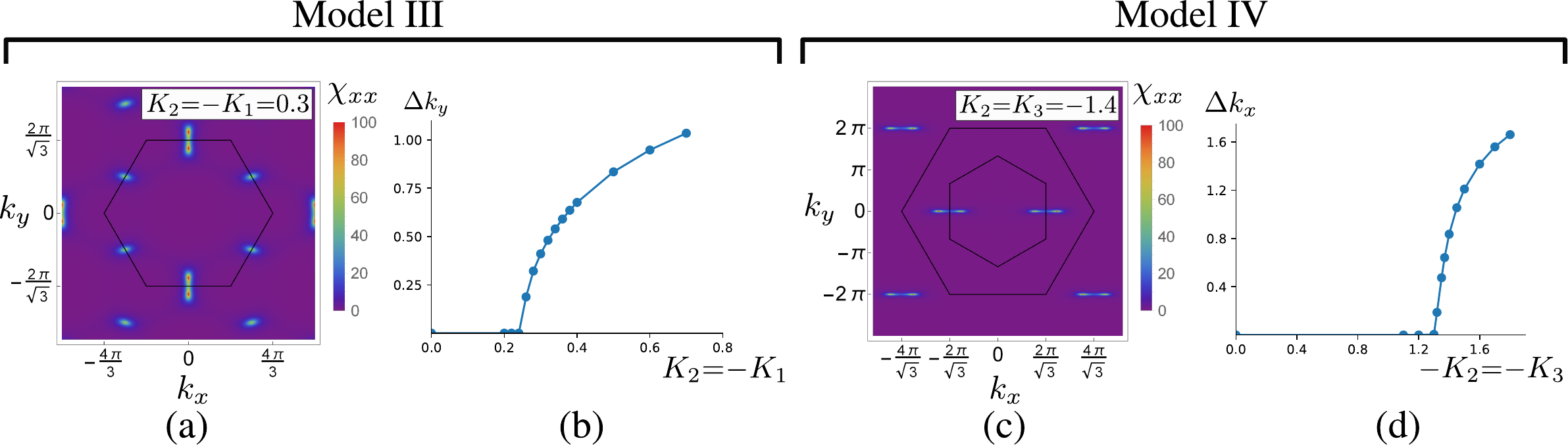}
\caption{Mean-field susceptibility $\chi^{xx}(\mathbf{k})$ and magnetic Bragg-peak evolution for models III and IV on the triangular and honeycomb lattices, respectively: (a) Susceptibility $\chi^{xx}(\mathbf{k})$ of model III at $K_2=-K_1=0.3$, (b) peak displacement $\Delta k_y$ relative to the tetrahedral order position (midpoints of the edges of the first Brillouin zone) as a function of $K_2=-K_1$ for model III, (c) susceptibility $\chi^{xx}(\mathbf{k})$ of model IV at $K_2=K_3=-1.4$, (d) peak displacement $\Delta k_x$ relative to the cubic order position (midpoints of the edges of the first Brillouin zone) as a function of $-K_2=-K_3$ for model IV. All data is collected right above the respective critical temperature $T_\text{c}$.}
\label{fig: TetraCubicMFT}
\end{figure*}

\subsubsection{Model I: 120$^\circ$-N\'eel order on the triangular lattice perturbed by $K_1$}
Here, we briefly revisit the well-known $\mathds{Z}_2$-vortex phase which is stabilized when perturbing the 120$^\circ$-N\'eel order on the triangular lattice by (positive or negative) nearest neighbor Kitaev couplings $K_1$~\cite{rousochatzakis16,becker15}. As can be seen in Fig.~\ref{fig: 120MFT}(a), showing a representative plot of the mean-field susceptibility $\chi^{xx}(\mathbf{k})$ at $K_1=-1.7$, the magnetic Bragg peaks, which reside at the corners of Brillouin zone in the case of pure 120$^\circ$-N\'eel order, have shifted along the boundaries of the Brillouin zone. Note that without loss of generality, we chose to depict the $xx$-component of the susceptibility. Due to the special spin/real-space symmetry of the Kitaev model [according to which the system remains invariant under a combined 120$^\circ$ real-space rotation and a 120$^\circ$ spin-space rotation around the $(1,1,1)$ axis] the $yy$ and $zz$-susceptibilities just correspond to 120$^\circ$-rotated versions of the $xx$-susceptibility. The peak displacement $\Delta k_x$ on the Brillouin-zone boundary, relative to the corner position, is shown in Fig.~\ref{fig: 120MFT}(b) as a function of $K_1$. The data indicates a continuous peak shift which already sets in at infinitesimally small $|K_1|$. This behavior qualitatively agrees with the numerical findings in Ref.~\onlinecite{rousochatzakis16}. As an obvious difference, however, the subleading satellite peaks from higher-harmonics which are numerically found in Ref.~\onlinecite{rousochatzakis16} are not resolved on a mean-field level.

\subsubsection{Model II: 120$^\circ$-N\'eel order on the triangular lattice perturbed by $K_2$}
Interestingly, the 120$^\circ$-N\'eel order on the triangular lattice hosts another vortex phase for antiferromagnetic $K_2>0$ which, to the best of our knowledge, has so far not been explored. As shown in Fig.~\ref{fig: 120MFT}(c) for $K_2=2$, the peaks exhibit a vertical displacement with respect to the corner position, i.e., they evolve along a direction perpendicular to the Brillouin zone boundary. Due to the momentum space periodicity of $\chi^{\mu\mu}(\mathbf{k})$ the migration of peaks along this perpendicular direction must be accompanied by a peak splitting. As illustrated by the white line in Fig.~\ref{fig: 120MFT}(c), for larger $K_2$ the peak trajectory starts bending back towards the edge of the Brillouin zone such that at a certain strength of $K_2$ they reside at the midpoints of the edges. Due to this property both components $k_x$ and $k_y$ of the magnetic peaks typically lie at incommensurate momenta. In Fig.~\ref{fig: 120MFT}(d) we plot the displacement $\Delta k=\sqrt{\Delta k_x^2+\Delta k_y^2}$ relative to the 120$^\circ$-N\'eel position. Most obviously, the peak shift sets in at a finite perturbation strength of $K_2\approx1.7$ and evolves continuously above this value. Note that a continuous peak shift is only observed in the presence of a finite $J_2<0$ interaction while in the absence of these couplings the peaks show a direct jump from the corner position to the midpoints of the Brillouin zone edges. This indicates the importance of $J_2$ interactions in stabilizing a vortex phase. As we will see in Sec.~\ref{im_results}, despite the common parent state, this vortex phase exhibits pronounced differences compared to model I such as the coexistence of different vortex types.

\subsubsection{Model III: Tetrahedral order on the triangular lattice perturbed by $K_1$ and $K_2$}
For the tetrahedral order, the magnetic Bragg peaks are located at the centers of the edges of the Brillouin zone. When perturbing the system along the aforementioned nine directions in the space of $(K_1,K_2,K_3)$ only the case $K_1<0$, $K_2=-K_1$, $K_3=0$ yields a continuous shift of magnetic Bragg peaks. We find that the peaks evolve along a straight line perpendicular to the Brillouin zone boundary, see Fig.~\ref{fig: TetraCubicMFT}(a) for $K_2=-K_1=0.3$. From the six initial peaks of the tetrahedral order, two exhibit increasing weight while the other four show decreasing intensity. Again, the peak migration sets in at finite Kitaev couplings $K_2=-K_1\approx0.25$ [see Fig.~\ref{fig: TetraCubicMFT}(b)] and evolves continuously for larger perturbations. Our more detailed analysis of this parameter regime in Sec.~\ref{im_results} reveals vortices with a remarkable counterrotating motion of spins on different sublattices.

\subsubsection{Model IV: Cubic order on the honeycomb lattice perturbed by $K_2$ and $K_3$}
Finally, we find that the cubic order on the honeycomb lattice (whose magnetic Bragg peaks are located at the centers of the edges of the first Brillouin zone) can be continuously deformed by ferromagnetic Kitaev couplings $K_2=K_3<0$. Very much similar to model III, the peaks move along a direction perpendicular to the Brillouin zone boundary [see Fig.~\ref{fig: TetraCubicMFT}(c) for $K_2=K_3=-1.4$] and the shift sets in at finite Kitaev couplings $K_2=K_3\approx-1.3$ [see Fig.~\ref{fig: TetraCubicMFT}(d)]. As we will see in Sec.~\ref{im_results}, the vortex phase of model IV can be considered as a doubled version of the vortex phase of model III, hence, explaining the similarities of the corresponding mean-field susceptibilities.

\subsection{Iterative minimization results}\label{im_results}
Due to the approximative nature of the mean-field approach and the limited information contained in momentum resolved susceptibilities we now present results of the iterative minimization technique which determines the actual ground-state spin pattern in real space. For each of the four models in Table~\ref{table:1} we discuss two aspects: The precise winding of spins in an individual vortex and their possible arrangement into a vortex lattice.

\subsubsection{Model I: 120$^\circ$-N\'eel order on the triangular lattice perturbed by $K_1$}

\begin{figure*}
\centering
\includegraphics[width=0.95\textwidth]{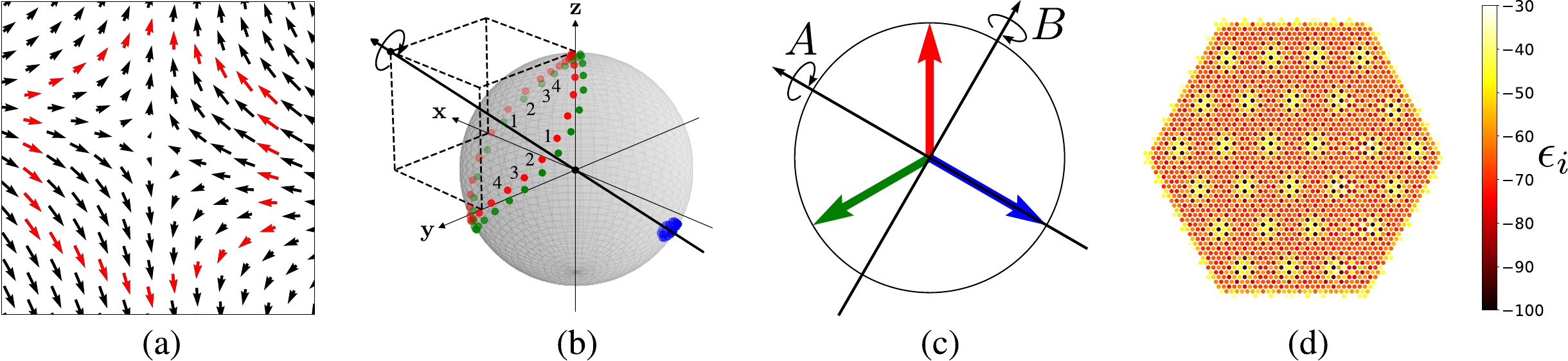}
\caption{Spin vortices for model I obtained with iterative minimization. (a) shows a representative $\mathds{Z}_2$ vortex at $K_1=-0.95$ where only one sublattice is depicted and the spins have been projected onto the $[1,1,1]$-plane. (b) Arrangement of spins on the Bloch sphere along the closed loop around the vortex core defined by the red arrows in (a). Different colors represent the different sublattices. To highlight the $(1,1,1)$-rotation axis (thick black arrow) we have added cubic edges in one octant (dashed lines). Spins with the same labels $1$, $2$, $3\ldots$ reside in the same unit cell. (c) Type-II vortices may feature different in-plane rotation axes of the local trio of spins: The depicted axis `A' coincides with the spin orientation in one of the three sublattice and is realized for all vortices of model I. The axis `B' is perpendicular to the spin orientation of one sublattice. This case is found for parts of the vortices of model II. (d) Local energies $\epsilon_i$ [see Eq.~(\ref{local_energy})] for the entire simulated system at $K_1=-1.6$.}
\label{fig: 120K1}
\end{figure*}
Iterative minimization identifies $\mathds{Z}_2$-vortices for sufficiently large Kitaev couplings $K_1\gtrsim 1$ and $K_1\lesssim-0.95$, i.e. when the vortex density is large enough such that their typical distance is smaller than the simulated system size. As an example, we illustrate a representative $\mathds{Z}_2$-vortex in Fig.~\ref{fig: 120K1}(a) where we project the spins onto the $[1,1,1]$-plane [this is the plane of rotation of the local trio of spins, see Fig.~\ref{fig: 120K1}(b)]. To better resolve the winding of the spins around the vortex core, we only show one sublattices of the 120$^\circ$-N\'eel order. As can be seen from the reduced length of the depicted spins in the center of the image, the largest out-of-plane components occur near the vortex core. The red arrows highlight the spins belonging to unit cells on a specified closed path around this core. For each unit cell along this path, Fig.~\ref{fig: 120K1}(b) shows the directions of the three spins (red, green, and blue) on the Bloch sphere where the red points correspond to the red arrows in Fig.~\ref{fig: 120K1}(a). One recognizes various characteristic vortex properties of this phase. Firstly, the spin directions on the Bloch sphere are concentrically arranged around the $(1,1,1)$-axis, indicating that the vortex is a result of a spin rotation within the $[1,1,1]$-plane. Secondly, the spins on two sublattices (red and green) perform a rotation when circling around the core, while the third sublattice remains fixed, pointing along the $(1,1,1)$-axis. Note that the two rotating sublattices have opposite spin directions when projected onto the $[1,1,1]$-plane. This is indicated by the labels $1$, $2$, $3\ldots$ where equal numbers correspond to spins belonging to the same unit cell [hence, in a plot of the type of Fig.~\ref{fig: 120K1}(a), the green sublattice would just show reversed arrows as compared to the red ones].

In Fig.~\ref{fig: 120K1}(c) we illustrate the spin arrangement in this vortex in a slightly different way by showing the local trio of spins together with their rotation axis (labelled `A') when going around the vortex core. Most crucially, this axis lies within the plane of the three spins indicative of a type-II vortex~\cite{kawamura84}. This is in contrast to a type-I vortex where the rotation axis is perpendicular to the plane of the 120$^\circ$-N\'eel parent state. It can also be seen that the rotation axis `A' is identical to the spin orientation in one sublattice which remains fixed along this path.

For larger values of the Kitaev coupling, vortices appeared arranged in a triangular superstructure. An example of this is shown in Fig.~\ref{fig: 120K1}(d) where we plot the local energies $\epsilon_i$ for each site $i$ defined by
\begin{equation}
\epsilon_i=\sum_\mu\sum_j J_{ij}^\mu S_i^\mu S_j^\mu\;.\label{local_energy}
\end{equation}
One can clearly see the vortex cores as defects in an otherwise homogeneous energy landscape. Note that the inclusion of a finite $J_2=-1$ in model I (see Table~\ref{table:1}) is justified by the empirical observation that a vortex lattice is much easier stabilized when a ferromagnetic second neighbor Heisenberg interaction is added. Without going into detail we would like to point out some further characteristic properties of this vortex lattice. For a more in-depth discussion we refer the interested reader to Ref.~\onlinecite{rousochatzakis16} where these properties have first been described. All vortices forming this superstructure are of type II with one sublattice remaining fixed. However, the rotation axis of the local trio of spins varies between different vortices but is always given by one of the four symmetry-equivalent $(1,1,1)$-directions [which are $(1,1,1)$, $(-1,-1,1)$, $(-1,1,-1)$, and $(1,-1,-1)$]. The rotation axis, hence, defines four subtypes of vortices which in the vortex superstructure form the same pattern as the four spin orientations of the tetrahedral state depicted in Fig.~\ref{fig1}(c)(ii). Together with the three possibilities for the fixed sublattice, there are twelve different vortex types and the magnetic unit cell comprises exactly one of each. In principle, there are twelve more vortex types which result from the former by inverting all spin directions (corresponding to their time-reversed counterparts). However, numerical outcomes never show these two sets of twelve vortex types being mixed up. In other words, depending on the initial spin configuration the system either exhibits vortices where the spins on the fixed sublattice only point in a $(1,1,1)$-type direction or vortices where the spins on the fixed sublattice only point in a $(-1,-1,-1)$-type direction. It is important to emphasize, however, that vortices with different fixed sublattices and/or rotation axes (including the distinction between type-I and type-II vortices) are all topologically equivalent since they can be transformed into each other by continuous spin rotations.

\begin{figure*}
\centering
\includegraphics[width=0.8\textwidth]{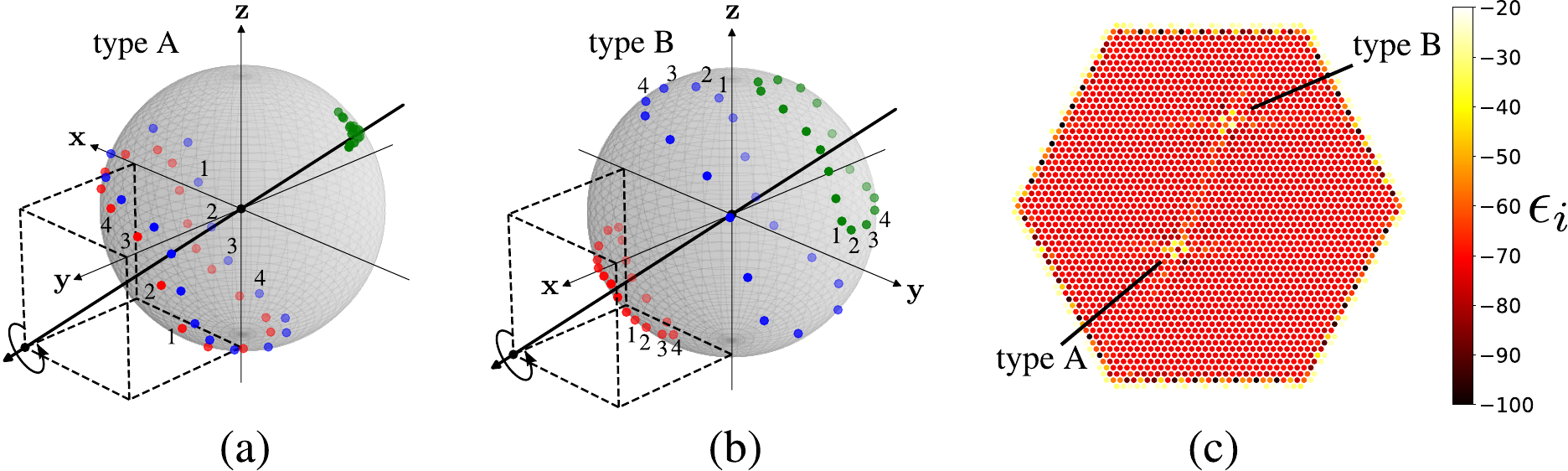}
\caption{Spin vortices for model II obtained with iterative minimization. (a), (b) Arrangements of spins on the Bloch sphere for a path around the vortex core in the case of type `A' and type `B' vortices, respectively [see Fig.~\ref{fig: 120K1}(c) for the corresponding rotation axes]. The vortices have been numerically obtained at coupling strengths of $K_2=1.75$ and $K_2=1.7$ respectively. Red, green, and blue points denote different sublattices of the 120$^\circ$-N\'eel state. To highlight the $(1,1,1)$-type rotation axis (thick black arrow) we have added cubic edges in one octant (dashed lines). (c) Local energies $\epsilon_i$ [see Eq.~(\ref{local_energy})] at $K_2=1.8$ for a numerical outcome including a type `A' and a type `B' vortex.}
\label{fig: 120K2}
\end{figure*}

\subsubsection{Model II: 120$^\circ$-N\'eel order on the triangular lattice perturbed by $K_2$}

Real space spin patterns generated with iterative minimization show vortices appearing at $K_2\approx 1.7$, in agreement with mean-field theory. The investigation of vortices in this system, however, turns out to be more difficult compared to model I. For example, in some outputs, vortices are grouped in tight clusters such that they become hard to distinguish. As discussed further below, this is likely due to a very complicated ground state which is hard to identify numerically. A closer analysis was performed on all isolated defects found in this model. Interestingly, two distinct types of vortices were detected. The first is qualitatively identical to the type II vortices found in model I, with the fixed sublattices aligned along one of the $(111)$ directions in spin space [see axis `A' in Fig.~\ref{fig: 120K1}(c)]. The second type of vortex involves rotations of all sublattices where the rotation axis is perpendicular to one of the three spin directions of the 120$^\circ$-N\'eel state [see axis `B' in Fig.~\ref{fig: 120K1}(c)] but is again given by one of the $(111)$ directions. Since the rotation axis lies within the plane of the 120$^\circ$-N\'eel order, these vortices are, likewise, of type II. Examples of vortices with these two distinct rotations are plotted in Fig.~\ref{fig: 120K2}(a) and (b), where spins along a path around the core are depicted on the Bloch sphere.

Furthermore, we show in Fig.~\ref{fig: 120K2}(c) the local energies $\epsilon_i$ of an output with one vortex of each type. Most notably, both vortices are energetically indistinguishable indicating that they may coexists in the real ground state. However, none of our simulations showed a vortex crystal such that we could not reveal the pattern in which these two vortex types arrange. We believe this is due to numerical difficulties in detecting a potentially very complex ground state. More precisely, the complications are twofold: Firstly, the total number of different vortex types in this system is vast. As explained before, type `A' vortices appear in 24 different species, and by the same argument one can arrive at the same number for vortices of type `B'. This results in a total of $2\cdot24=48$ different vortex types which may altogether form a large magnetic unit cell. Secondly, different $K_2$ parameter regimes exhibit distinct numerical challenges such that an optimal strength for $K_2$, where simulations become feasible, might not exist. For $K_2\gtrsim 1.7$, i.e., close to the transition point, vortex densities are small such that the system only gains little energy by realizing a regular superstructure. These condensation energies may be well below our numerical accuracies. For larger $K_2$ our mean-field results indicate that the trajectory of Bragg peaks bends away from the initial high-symmetry line $k_x=\pm 2\pi/3$ [see Fig.~\ref{fig: 120MFT}(c)] such that both wave vector components $k_x$, $k_y$ become incommensurate. In this case, the lattice vectors of the magnetic unit cell will be twisted against the lattice vectors of the underlying triangular lattice with an irrational rotation angle between them (one may contrast this situation with the vortex phase of model I where the component $k_y=\pm 2\pi/\sqrt{3}$ remains commensurate and the orientations of the vortex and spin lattices agree). This adds another source of incommensurability to the system which is hard to capture in our finite simulated lattices. We, hence, keep a more detailed characterization of this phase for future studies.

\subsubsection{Model III: Tetrahedral order on the triangular lattice perturbed by $K_1$ and $K_2$}

\begin{figure*}
\centering
\includegraphics[width=0.7\textwidth]{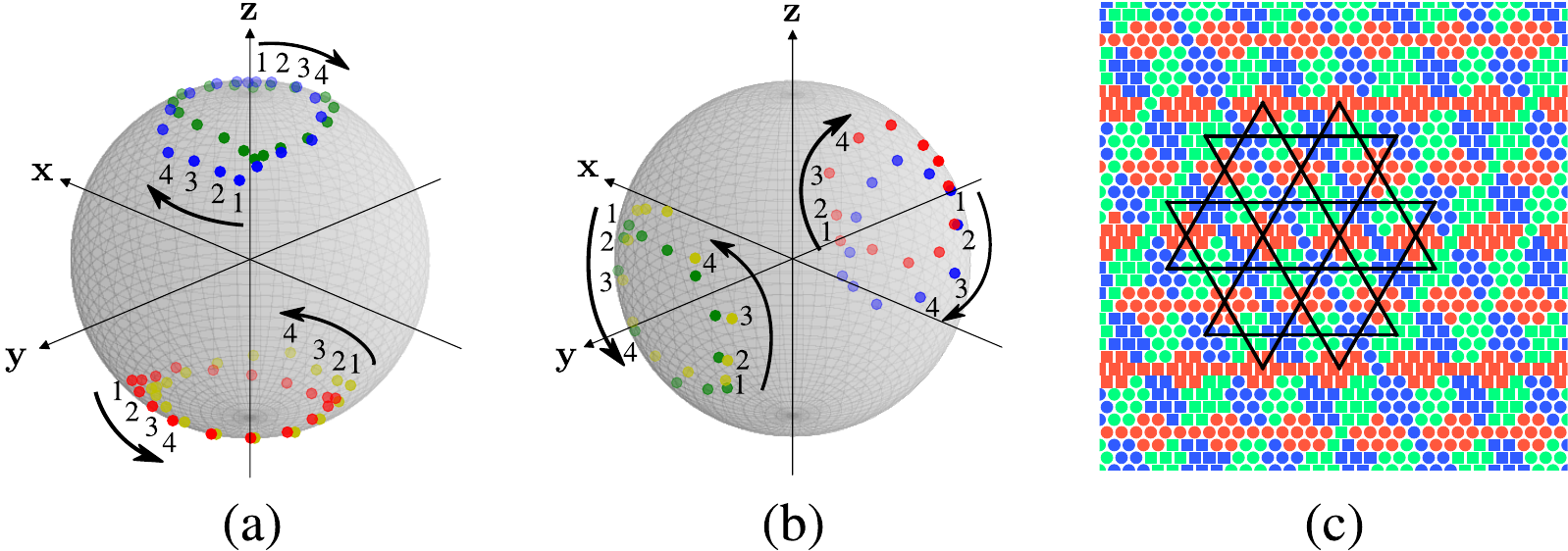}
\caption{Spin vortices for model III from iterative minimization. (a), (b) Spin arrangements of the four tetrahedral sublattices (colored red, green, blue, yellow) for closed paths surrounding vortex cores obtained for $K_2=-K_1=0.28$. The two vortices in (a) and (b) show spin rotations around the $z$ and $y$ axes, respectively. Note that for each vortex the sublattices break up into pairs where one pair exhibits a clockwise and the other a counterclockwise rotation, see text for details. (c) Real-space spin configuration obtained with iterative minimization for $K_2=-K_1=0.5$ where each site is colored according to the spin component $\mu\in\{x,y,z\}$ with the largest absolute value $|S_i^\mu|$ (red, green, blue correspond to $x$, $y$, $z$, respectively). Square (circular) symbols indicate that for the first, i.e. red, sublattice the corresponding signed spin component $S_i^\mu$ is positive (negative). We draw a kagome lattice on top of the configuration to highlight the formation of a superstructure.}
\label{fig: TetraK1K2}
\end{figure*}
This model exhibits ferromagnetic first and antiferromagnetic second neighbor Kitaev couplings with the same absolute value, i.e., $K_2=-K_1>0$. In good agreement with mean-field theory, numerical data from iterative minimization shows vortices for $K_2=-K_1>0.27$. In Figs.~\ref{fig: TetraK1K2}(a) and (b) we show the spin directions on the Bloch sphere along loops around the cores for two different vortices. Note that we distinguish here between the four sublattices of the tetrahedral-type states which are colored red, green, blue, and yellow. Interestingly, the detected spin patterns are in striking contrast to the previous models: All vortices feature rotation axes along one of the cartesian $x$, $y$ or $z$ spin directions. From the perspective of the underlying parent state, these axes are of the same type as the gray line in Fig.~\ref{fig1}(c)(i) connecting the midpoints of two opposite tetrahedral edges. Consequently, the winding of vortices is described by variations of the angle $\varphi$. The four tetrahedral spin directions, however, do not rotate as one entity with their relative orientations fixed, but rather show a splitting into pairs of sublattices where one pair rotates with an angle $\varphi$ while the other pair features a reversed motion with an angle $-\varphi$. This is indicated in Figs.~\ref{fig: TetraK1K2}(a) and (b) by the numbers $1$, $2$, $\ldots$ labelling sites within the same unit cell. Furthermore, the opening angle $\alpha$ between the spins in each pair is approximately proportional to the distance from the vortex core such that in the center the spins are closely aligned (not shown in Fig.~\ref{fig: TetraK1K2}). Hence, at $\mathbf{r}=(r,\phi)$ describing a point on the lattice in polar coordinates relative to the core, the local tetrahedral spin configuration may be approximated by
\begin{equation}
\varphi(\phi)=\pm\phi+\phi_0\;,\;\;\alpha(r)=ar\;,
\end{equation}
where ``$\pm$'' refers to the two pairs, $\phi_0$ is an angular offset and $a$ is a proportionality constant. In other words, the local spin patterns in the vicinity of vortex cores explore parts of the degenerate tetrahedral manifold. Since away from the vortex core, each rotating pair of spins is non-collinear and hence, spans a local SO(3) configuration space, these vortices may be classified as being of $\mathds{Z}_2$ type.

We further find that for all vortices the pairing of sublattices is directly tied to the cartesian rotation axis in spin space, in a way that matches the direction of Kitaev anisotropy on different nearest neighbor bonds: For a vortex with rotation axis $z$ the sublattices break up into pairs (blue, green) and (yellow, red) which are exactly those pairs of sites carrying $z$-type Kitaev interactions $S_i^z S_j^z$, and equivalently for the other rotation axes, see Fig.~\ref{fig1}(c)(ii). Hence, from their sublattice structure and rotation axes one may distinguish between three different vortex types. For each of these three species there are two subtypes of vortices depending on which of the two pairs points along the positive (negative) cartesian rotation axes (up to the sense of rotation these two vortex types are time-reversal partners of each other). For example, for $z$-type vortices one can distinguish between cases where the blue and green sublattices have positive spin components along the $z$-axis and cases where these components are negative (the yellow and red sublattices feature opposite signs of the $z$-components in both cases). In total, this results in six different vortex types. It may seem surprising that model III permits a much smaller number of vortex types compared to the previous models, even though the underlying tetrahedral parent state appears more complicated. The reason for this is the aforementioned spin-space/real-space locking of model III where the sublattice pairing is tied to the rotation axis. This is in contrast to model I where the fixed sublattice of a vortex is independent of its rotation axes, hence, leading to a large number of vortex types.

Note that the spin-orbit symmetry of the Kitaev model [according to which the system remains invariant under a 120$^\circ$-rotation in real space, combined with a 120$^\circ$ rotation around the $(1,1,1)$ axis in spin space, swapping $xx\rightarrow yy\rightarrow zz \rightarrow xx$] transforms vortex types with different cartesian rotation axes into each other. Therefore, assuming that this symmetry is not spontaneously broken on a global level, one expects that vortices with all three rotation axes coexist in the system. Above $K_2=-K_1\approx0.4$ where the vortices become dense enough such that we could resolve a vortex lattice, this is indeed observed. In Fig.~\ref{fig: TetraK1K2}(c) we show the output of a simulation where we color each site according to the spin component with the largest absolute value $|S_i^\mu|$ where $\mu\in\{x,y,z\}$. Due to the fact that in the center of each vortex, the spins are nearly aligned with the respective cartesian axis we can easily distinguish between the different vortex types. As can be seen, the vortices form a kagome lattice, where each `sublattice' of the kagome superstructure hosts vortices with the same cartesian rotation axes. Additionally, the aforementioned two subtypes of vortices (which differ by the sign of the projection onto the rotation axis) are also present in the system. Particularly, for vortices with a given rotation axis the two subtypes form alternating stripes as indicated by circular and square symbols in Fig.~\ref{fig: TetraK1K2}(c).

\subsubsection{Model IV: Cubic order on the honeycomb lattice perturbed by $K_2$ and $K_3$}
\begin{figure*}
\centering
\includegraphics[width=0.7\textwidth]{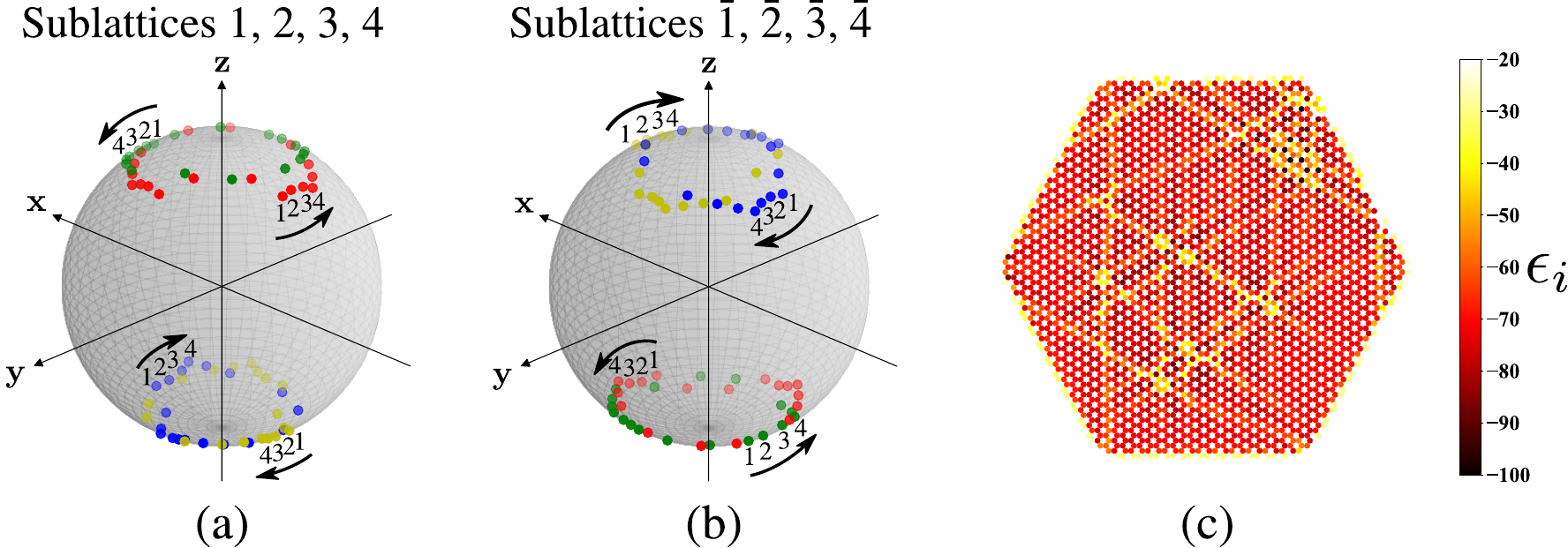}
\caption{Spin vortices for model IV from iterative minimization. (a), (b) Bloch sphere illustration of a single vortex at $-K_2=-K_3=1.41$ with a cubic ordered parent state where (a) shows the sublattices $1$, $2$, $3$, $4$ while (b) depicts the sublattices $\bar{1}$, $\bar{2}$, $\bar{3}$, $\bar{4}$, colored red, green, blue, yellow in each case [see Fig.~\ref{fig2}(b)(i) for the definition of sublattices]. The vortex properties in (a) and (b) are identical to those of model III. (c) Local sites energies for a numerical outcome at $-K_2=-K_3=1.5$ showing an irregular pattern of defects.}
\label{fig: CubicK2K3}
\end{figure*}
Due to the bipartite nature of the honeycomb lattice which consists of two interpenetrating triangular lattices, it is clear that a vortex phase similar to the one of model III can also be stabilized in a honeycomb model. In the simplest case this amounts to implementing model III on both triangular sublattices and only allowing for small couplings between them (implying small $J_1$, $J_3$, $K_1$, $K_3$, $\ldots$). The second neighbor triangular lattice couplings $J_2$ and $K_2$ which were needed to generate the vortex phase in model III would then correspond to fifth neighbor couplings $J_5$, $K_5$ on the honeycomb lattice. Here, we show that a duplicated version of the vortex phase of model III may already be stabilized in a simpler system with honeycomb interactions only ranging up to third neighbors. Interestingly, this phase even occurs for sizable inter-sublattice couplings $J_1$, $J_3$, and $K_3$, see Table~\ref{table:1}. As discussed in Sec.~\ref{cubic}, the two Heisenberg couplings $J_1$ and $J_3$ are actually needed for obtaining a cubic ordered parent state.

In agreement with our mean-field analysis, iterative minimization finds vortices for $-K_2=-K_3>1.4$. An example is shown in Figs.~\ref{fig: CubicK2K3}(a) and (b) where we plot the spin configurations of a single vortex on the Bloch sphere. For better illustration, we split up the eight sublattices of the cubic order into groups of four, shown in subfigures (a) and (b), respectively, where each group represents one of the two triangular sublattices of the honeycomb lattice. Using the convention of Fig.~\ref{fig2}(b)(i), subfigure (a) shows sublattices $1$, $2$, $3$, $4$ while (b) depicts sublattices $\bar{1}$, $\bar{2}$, $\bar{3}$, $\bar{4}$. Furthermore, pairs of data points in (a) and (b) with the same color correspond to pairs $(\alpha,\bar{\alpha})$ where $\alpha\in\{1,2,3,4\}$. Considering Figs.~\ref{fig: CubicK2K3}(a) and (b) separately, it can be seen that the properties from model III directly carry over: Vortex rotations always occur around cartesian axes (here, only a vortex with a $z$ rotation axes is shown) and the four sublattices in each plot split up into pairs showing a counterrotating motion. When comparing the two plots, one further finds that pairs of spins $(\alpha,\bar{\alpha})$ in the same unit cell have opposite directions which agrees with the spin pattern of the cubic parent state discussed in Sec.~\ref{cubic}.

Due to these properties one would expect that similar to model III the system hosts six vortex types which condense into a kagome superstructure. However, our numerical data never shows regular vortex lattices. A typical outcome at $-K_2=-K_3=1.5$ is plotted in Figs.~\ref{fig: CubicK2K3}(c) where the site energies $\epsilon_i$ clearly indicate local defects but without arranging in a regular pattern. We speculate that this might be due to the increased unit cell of the honeycomb lattice and/or due to the more complicated underlying spin model with interactions ranging up to third neighbors.

\section{Conclusions} \label{DandC}

In this work we have have studied the generation of $\mathds{Z}_2$-vortex phases and vortex crystals in Kitaev-Heisenberg models, in various geometries and parent orders beyond the 120$^\circ$ N\'eel state. We have probed large classes of systems, following a two step approach: Using an analytical mean field method, we first searched for the characteristic peak shifting in reciprocal space which allowed us to reduce the number of systems to a few candidate models. These remaining models have then been treated with the iterative minimization technique to study their ground state spin configurations in real space.

We have identified and discussed four different vortex phases where two of them are based on the 120$^\circ$ N\'eel state (models I and II) while the other two rely on the tetrahedral order or variants thereof (models III and IV). Note that model I is identical to the system studied in Ref.~\onlinecite{rousochatzakis16}. One of our main findings is that these two groups of models show striking differences in the nature of their vortices. The two systems with 120$^\circ$ N\'eel parent order host type II vortices where the winding of the planar tripods of spins around the vortex cores features an in-plane rotation axis. This axis points along one of the diagonal $(1,1,1)$-type directions in spin space, revealing a spin-locking mechanism which is typical for many Kitaev systems~\cite{price13}. While in model I the rotation axis is oriented such that the spin directions in one sublattice remain fixed in the vicinity of a vortex core, model II also allows for vortices where all three sublattices show a rotation.

Models III and IV feature distinctly different vortex properties which are rooted in the fact that their tetrahedral and cubic parent orders exhibit a continuous degeneracy (that may be parametrized by two angles). In contrast to the previous models where the local tripods of spins rotate like a rigid body, these systems show vortices where the sublattices of the tetrahedral/cubic orders split up into two groups which rotate around a common axis but with opposite sense of rotation. We, hence, conclude that by exploring parts of the degenerate manifold of states, these vortices gain energy compared to a `rigid body rotation'. Another difference to the previous systems is that the special rotation axes are given by the cubic $(1,0,0)$-type directions.

Vortices with parent orders beyond the 120$^\circ$ N\'eel state open up various interesting future directions of research. One may, for example, search for vortices from tetrahedral order in the pure parent $J_1$-$J_2$ Heisenberg model on the triangular lattice, i.e., without adding Kitaev interactions. This model possibly exhibits similar phenomena as the nearest neighbor triangular Heisenberg model where $\mathds{Z}_2$-vortices from 120$^\circ$ N\'eel order are stabilized by thermal fluctuations and undergo a BKT-like vortex binding-unbinding transition at finite temperatures~\cite{kawamura84}. Furthermore, one may try to stabilize a vortex phase where the non-planar local spin arrangements show a `rigid body rotation' around the vortex cores. In models III and IV, this might become possible when adding further types of anisotropic interactions, such as Dzyaloshinskii-Moriya couplings or $\Gamma$-exchange, which have not been considered here. An alternative would be to try to generate such phases based on the twelve-sublattice cuboc 1 or cuboc 2 parent orders on the kagome lattice. These states are (up to global rotations) non-degenerate in the classical kagome Heisenberg model and, hence, do not permit the sublattice-splitting mechanism of models III and IV. We have already started to search for such phases, however, at least for Kitaev interactions up to third neighbors, the cuboc orders were never seen to evolve into vortex phases. It would still be worth adding Dzyaloshinskii-Moriya and/or $\Gamma$-interactions which may potentially stabilize novel and unexplored vortex phases.

\section{Acknowledgements}
We thank L. Messio, S. Trebst and M. Daghofer for helpful discussions.
\bibliography{ref}

\end{document}